\documentclass[twocolumn,aps,american,pra,floatfix,nofootinbib,superscriptaddress,longbibliography]{revtex4-2}
\usepackage[unicode=true,pdfusetitle, bookmarks=true,bookmarksnumbered=false,bookmarksopen=false, breaklinks=false,pdfborder={0 0 0},backref=false,colorlinks=false]{hyperref}
\hypersetup{colorlinks,linkcolor=blue,citecolor=red,urlcolor=cyan}
\usepackage{graphics,epstopdf,amsthm,amsmath,amssymb,mathptmx,colortbl,color,bm,framed,mathrsfs}
\usepackage[T1]{fontenc}
\usepackage[braket, qm]{qcircuit}
\usepackage{commath}
\usepackage[capitalise]{cleveref}
\usepackage{multirow}
\makeatother

\newcommand{\bes} {\begin{subequations}}
\newcommand{\ees} {\end{subequations}}
\newcommand{\beq}{\begin{equation}}
\newcommand{\eeq}{\end{equation}}

\newcommand{\BV}{\text{BV}}

\usepackage[demo]{graphicx}

\usepackage{mwe}
\usepackage[export]{adjustbox}
\usepackage[normalem]{ulem}
\begin{document}
\title{Demonstration of algorithmic quantum speedup}

\author{Bibek Pokharel}
\email{pokharel@usc.edu}
\affiliation{Department of Physics \& Astronomy}
\affiliation{Center for Quantum Information Science \& Technology}

\author{Daniel A. Lidar}
\email{lidar@usc.edu}
\affiliation{Department of Physics \& Astronomy}
\affiliation{Center for Quantum Information Science \& Technology}
\affiliation{Department of Electrical \& Computer Engineering}
\affiliation{Department of Chemistry\\
University of Southern California, Los Angeles, CA 90089, USA}

\date{\today}

\maketitle

\textbf{Quantum algorithms theoretically outperform classical algorithms in solving problems of increasing size~\cite{Deutsch:92,bernsteinQuantumComplexityTheory1997,groverFastQuantumMechanical1996,shorPolynomialTimeAlgorithmsPrime1999,montanaroQuantumAlgorithmsOverview2016}, but computational errors must be kept to a minimum to realize this potential.  Despite the development of increasingly capable quantum computers (QCs), an experimental demonstration of a provable algorithmic quantum speedup~\cite{ronnowDefiningDetectingQuantum2014} employing today's non-fault-tolerant, noisy intermediate-scale quantum (NISQ)~\cite{Preskill:2018aa} devices has remained elusive. Here, we unequivocally demonstrate such a speedup, quantified in terms of the scaling with the problem size of the time-to-solution metric. We implement the single-shot Bernstein-Vazirani algorithm, which solves the problem of identifying a hidden bitstring that changes after every oracle query, utilizing two different $27$-qubit IBM Quantum (IBMQ)~\cite{IBMQuantum2022} superconducting processors. The speedup is observed on only one of the two QCs (ibmq\_montreal) when the quantum computation is protected by dynamical decoupling (DD) -- a carefully designed sequence of pulses applied to the QC that suppresses its interaction with the environment~\cite{Viola:98,Suter:2016aa}, but not without DD. In contrast to recent quantum supremacy demonstrations~\cite{Arute:2019aa,wu2021strong,Zhong:2021wv}, the quantum speedup reported here does not rely on any additional assumptions or complexity-theoretic conjectures~\cite{Aaronson:2016aa,Harrow:2017aa,Hangleiter:22} and solves a bona fide computational problem, in the setting of a game with an oracle and a verifier.}\\

\noindent\textbf{Main}\\
\noindent The quest to demonstrate a quantum speedup for a computational problem, i.e., an algorithmic quantum speedup, has motivated the field of quantum computing from its inception. Early theoretical breakthroughs, most notably Shor's factorization algorithm~\cite{shorPolynomialTimeAlgorithmsPrime1999}, set in motion a remarkable surge of quantum hardware development aimed at realizing the potential of a quantum speedup. The last several years have witnessed the deployment of both academic and commercial quantum computing platforms accessible to experimentation via the cloud. Better-than-classical algorithmic performance has been demonstrated a number of times, e.g., on ion-trap~\cite{figgattComplete3QubitGrover2017,wrightBenchmarking11qubitQuantum2019},  superconducting~\cite{royProgrammableSuperconductingProcessor2020,Huang:2021}, photonic~\cite{saggioExperimentalQuantumSpeedup2021,centroneExperimentalDemonstrationQuantum2021,xiaQuantumEnhancedDataClassification2021,zhouExperimentalQuantumAdvantage2022}, and Rydberg atom~\cite{Ebadi:22} quantum processors, in most cases by exceeding the corresponding classical algorithmic success probability at a fixed problem size, or a small set of such sizes. 
However, as we explain in more detail below, this does not yet amount to an algorithmic quantum speedup demonstration, which involves the more stringent criterion of the scaling with problem size of the time-to-solution metric. 
We provide a comprehensive survey of existing experimental demonstrations of better-than-classical algorithmic results in \cref{app:A}.

To qualify as a genuine, \emph{provable} algorithmic quantum speedup, we insist that the speedup be relative to a problem whose classical solution cannot be improved in principle and is unconditional over the range of accessible problem sizes. In other words, the scaling advantage of the time-to-solution does not rely on any assumptions and persists up to the largest number of qubits compatible with the QC being used to demonstrate it.
To satisfy this stringent notion of an algorithmic quantum speedup, we revisit the Bernstein-Vazirani (BV) algorithm, which was one of the very first theoretical examples of a quantum speedup and quantum/classical complexity class separation~\cite{bernsteinQuantumComplexityTheory1997}.
In the original BV problem, an oracle outputs $f_b(x) = b \cdot x \ (\text{mod }2)$, where $x$ and $b$ are both length-$n$ bitstrings. Here $x$ is a guess provided by the user and $b$ is a secret bitstring the user is trying to learn in as few oracle queries as possible. The best classical algorithm requires $n$ queries of the type $f_b(0\dots 01_i 0\dots 0)$, $i\in\{1,2,\dots,n\}$, each of which returns one unknown bit $b_i\in b$.
By solving the problem with certainty in a \emph{single} query, the BV algorithm provides a linear speedup over the best-classical algorithm. 

\begin{figure}[t]
\centering
\vspace{-5mm}
    \begin{displaymath}
\large
\Qcircuit @C=1em @R=0.9em @!R {
 & & &  \mbox{\ \ \ \ \ BV Oracle} &  &  \\
\lstick{| 0 \rangle} & \gate{H} & \ctrl{5} & \qw & \qw & \qw & \gate{H} & \meter  \\
\lstick{| 0 \rangle} & \gate{H} & \qw & \ctrl{4} & \qw & \qw & \gate{H} & \meter  \\
& \vdots &  & & & & \vdots & \vdots  \\
\lstick{| 0 \rangle}  & \gate{H} & \qw & \qw & \ctrl{2} & \qw & \gate{H} & \meter  \\
\lstick{| 0 \rangle} & \gate{H} & \qw & \qw & \qw & \ctrl{1} & \gate{H} & \meter  \\
\lstick{| 1 \rangle}  & \gate{H}  & \targ & \targ & \targ & \targ & \gate{H} & \meter \gategroup{2}{3}{7}{6}{0.6em}{-}
}
    \end{displaymath}
\includegraphics[width=\columnwidth, valign=t]{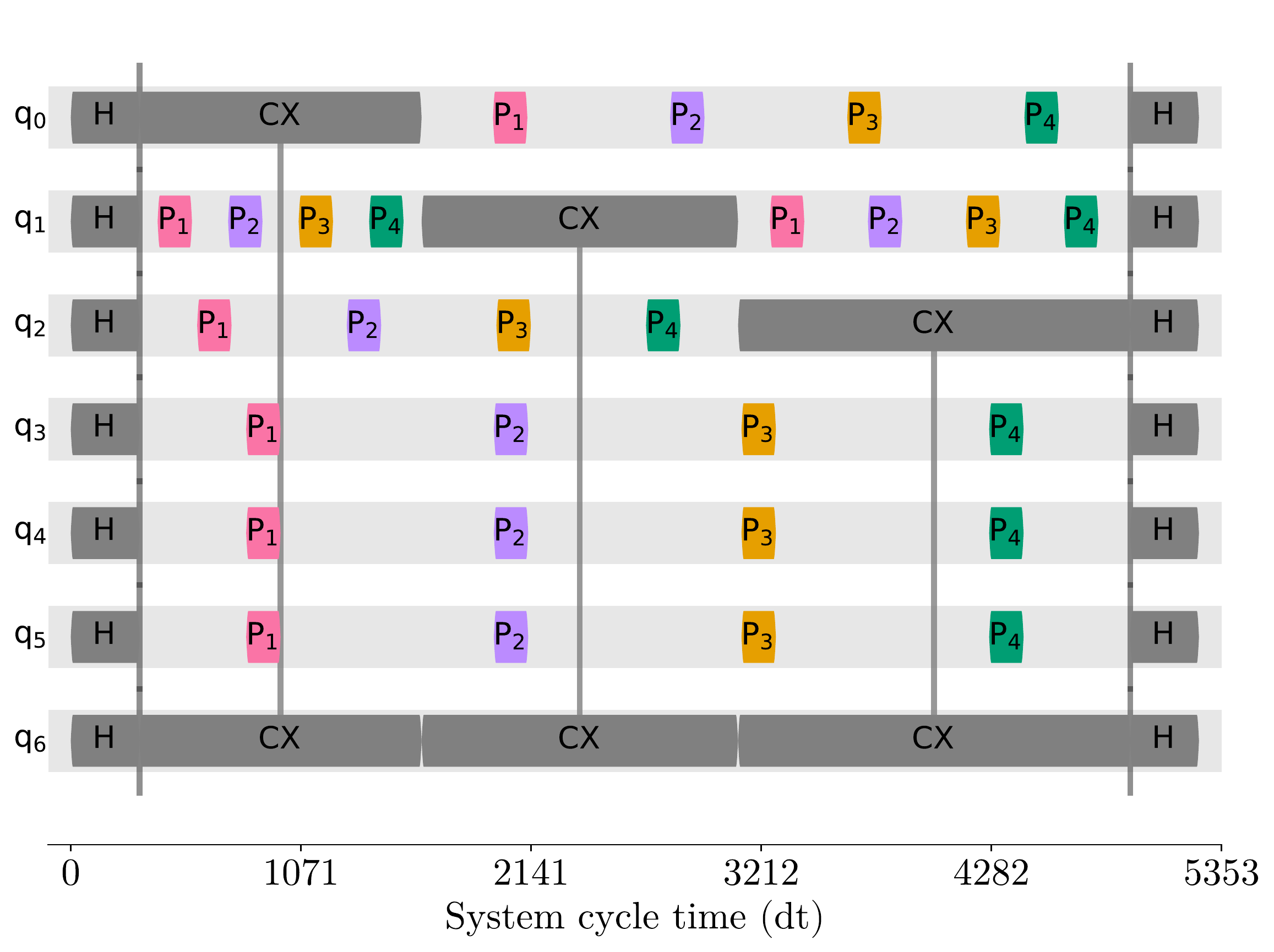} 
\caption{Top: Circuit for the BV algorithm with the unknown bitstring $b=1\dots1$. In the BV guessing game (or ssBV-$n$ problem), the oracle changes after each circuit execution and encodes a new bitstring $b$. Note that the internal structure of the circuit is hidden from the player, so the fact that the BV circuit can be efficiently simulated classically (since it uses only Clifford gates~\cite{Bravyi:2016wf}) is immaterial. Each BV-$n$ circuit requires $n+1$ qubits and a circuit of depth $\ge n+3$ (with equality only for fully connected architectures). A controlled-NOT (CNOT) or identity gate is performed from qubit $i$ to the ancilla qubit if $b_i= 1$ or $0$, respectively.
Bottom: The same circuit for BV-$6$ with DD pulses during idle times; $ P_i$ denotes pulses. Pulse placement is schematic but illustrates all the principles we used in practice: (i) DD fills all available idle spaces, (ii) pulse intervals are varied depending on the available idle time per qubit, (iii) pulse types can differ (i.e., P$_i \neq$ P$_j$), (iv) a single qubit experiences multiple repetitions of a sequence if there are disjoint intervals (as in the case of q$_1$). The actual timeline is shown, in units of  $dt = 2/9 \text{ ns}$ -- the inverse sampling rate of the backend's arbitrary waveform generators.
}
\label{fig:bv_circuit}
\end{figure}

Here we consider a modified, single-shot version of BV,
% (also called probabilistic BV~\cite{naseriEntanglementCoherenceBernsteinVazirani2022}), 
denoted ssBV-$n$, \emph{where the hidden bitstring $b$ changes after every query} [see~\cref{fig:bv_circuit}(top)]. We colloquially refer to this as the ``BV guessing game'': after one query of the single-shot oracle, the player is allowed one guess of the bitstring $b$. If the verifier confirms that the guess is correct, the player wins; if the guess is wrong, the game continues with a new oracle encoding a new bitstring. 

In this setting, the best classical algorithm is to each time query the oracle with bitstring $0\dots 01_i 0\dots 0$ ($i$ is arbitrary), which reveals $b_i$, and then guess the remaining $n-1$ bits. This yields classical success probability $p_{s} = 2^{1-n}$, only twice better than a random guess. In stark contrast, a player with access to a QC running the original BV algorithm has success probability $p_{s}=1$ after each query, which becomes an exponential advantage in the speedup ratio (defined below) over the classical setting. \\ 

\noindent\textbf{Quantum speedup quantified}\\
\noindent In a head-to-head comparison of success probabilities, $p_s > 2^{1-n_0}$, for a fixed problem size $n_0$,  implies a better-than-classical result. This is the context in which better-than-classical results have been achieved for the Grover and BV algorithms~\cite{wrightBenchmarking11qubitQuantum2019,figgattComplete3QubitGrover2017, royProgrammableSuperconductingProcessor2020}.
However, the success probability at fixed problem size is not a reliable measure of quantum speedup, as detecting an algorithmic speedup requires the scaling with problem size. Moreover, $p_s$ is itself a function of the time $t_r(n)$ taken to run the calculation, i.e., the time required to run the complete quantum or classical circuit once. Instead, we quantify quantum speedup in terms of the scaling with the problem size $n$ of the speedup ratio of the classical and quantum total runtimes: $S(n)=\frac{\text{TTS}_C(n)}{\text{TTS}_Q(n)}$, where the total runtime is quantified using the well-established time-to-solution (TTS) metric~\cite{ronnowDefiningDetectingQuantum2014}:
\begin{equation}
    \text{TTS}(n) =  t_r(n) R(n)\ , \quad R(n) = \frac{\log (1-p_d)}{\log (1-p_s(t_r(n)))}.
\end{equation}
$\lceil R(n) \rceil$ is the number of repetitions -- oracle calls in the present context -- needed to find a solution at least once with desired probability $p_d$, given that a single repetition succeeds with probability $p_s(t_r(n))$; we set $p_d=0.99$ henceforth. 
Thus, the TTS quantifies the total time it takes to win the BV guessing game, whether classically or with access to a QC.

%Here, we focus on the pure runtime, which includes
We choose to measure $t_r(n)$ in terms of 
the circuit execution time and the readout duration, and ignore precompilation and postprocessing overheads, as the latter are inherently classical.
It follows from the BV circuit structure (\cref{fig:bv_circuit}) that $t_r(n)= c \tau_{2q} n + \tau_0$, where $1\leq c \leq 2$ depends on the qubit connectivity graph, with the two limits corresponding to all-to-all connectivity ($c=1$) and a chain ($c=2$). For our IBMQ implementation, we found $c \approx 1.76$ (see \cref{app:D} for more details).  The two-qubit gate time, $\tau_{2q}$, and the sum of the single qubit and readout times, $\tau_0$, depend on the specific QC and can differ by orders of magnitude across platforms; e.g., $\tau_{2q}$ is measured in microseconds for trapped-ion qubits and nanoseconds for superconducting qubits. 

Let $n_{\max}$ denote the largest number of data qubits available to a quantum algorithm; e.g., in the BV algorithm case, which requires one ancilla qubit, $n_{\max} = n_{\text{tot}}-1$, where $n_{\text{tot}}$ is the total number of programmable physical qubits available on the physical quantum device. 
Any scaling of $S(n)$ that is extracted from a QC with a relatively small $n_{\text{tot}}$, as is invariably the case in the current NISQ era~\cite{Preskill:2018aa}, is naturally subject to finite-size effects. Thus, the best one can hope for is that extrapolations to $n>n_{\max}$ are meaningful, and any conclusions based on such extrapolations must be revisited when devices with larger $n_{\text{tot}}$ become available. With this in mind, we estimate the scaling of the speedup ratio $S(n)$ by computing the most conservative estimate allowed by the experimental data, as explained below. This extrapolated scaling can be used to compare different QCs. 

In the ssBV-$n$ case, we expect $\text{TTS}_Q(n)$ to scale as $n 2^{\lambda n}$ (with $\lambda >0$) due to decoherence, instead of as $t_r \sim n$, as would be the case for a noiseless QC. When ssBV-$n$ is solved classically, computing $f_b(x) = b \cdot x \ (\text{mod }2)$ also takes time $\propto n$ (the cost of adding $n$ bits), so we obtain $\text{TTS}_C(n) \propto n/\log_2 (1-2^{1-n}) \approx n 2^{n-1}$. We thus expect 
\beq
S(n) \sim 2^{(1-\lambda) n}, \quad n \in[n_{\min}, n_{\max}] ,
\eeq
where $n_{\min}$ is identified empirically by excluding small-size effects.
\emph{We will declare a quantum speedup if the speedup exponent $\lambda <1$}. It is important to emphasize that the speedup exponent must be extracted using $n$ reaching up to and including $n_{\max}$, since otherwise one cannot hope to draw conclusions that reflect asymptotic scaling behavior.
Using this criterion, we demonstrate below that a statistically significant quantum speedup is achieved for DD-protected ssBV-$n$ quantum circuits, but no speedup is obtained for ``bare'' quantum circuits implemented without DD.\\

\noindent\textbf{Dynamical decoupling}\\
\noindent DD-protection has a long history of experimental demonstrations on various quantum devices (see Ref.~\cite{Suter:2016aa} for a review), and has also been shown to improve various performance metrics, such as qubit memory fidelity~\cite{Pokharel2018, souzaProcessTomographyRobust2020}, crosstalk mitigation~\cite{tripathiSuppressionCrosstalkSuperconducting2022}, quantum volume~\cite{jurcevicDemonstrationQuantumVolume2021}, and 
algorithmic fidelity~\cite{raviVAQEMVariationalApproach2021}. Various theoretical strategies for how to combine DD with quantum computation have been described~\cite{Viola:99a,KhodjastehLidar:08,khodjasteh:080501}, but we are unaware of prior experimental demonstrations of the use of DD to directly improve quantum algorithmic scaling.

We employ a ``decouple then compute'' strategy~\cite{West:10,Ng:2011dn}, whereby DD control pulses constituting short but complete DD sequences are interleaved with the quantum circuit by exploiting intervals when individual qubits in the corresponding quantum circuits are idle; see \cref{fig:bv_circuit}(bottom).  

A major challenge in using DD is that pulse imperfections can significantly deteriorate performance, necessitating a careful choice of DD sequence. Building on a survey of numerous known sequences~\cite{Ezzell:22}, we selected the universally robust (UR) sequence family~\cite{genovArbitrarilyAccuratePulse2017} as the top performance enhancer. This sequence was designed to suppress flip-angle errors and has been shown previously to enhance performance in superconducting-qubit-based NISQ devices~\cite{souzaProcessTomographyRobust2020,gautamProtectionNoisyMultipartite2021}. \\

\begin{figure}[t]
\centering
\includegraphics[trim = 5 0 0 0, clip, width=1.05\columnwidth, valign=t]{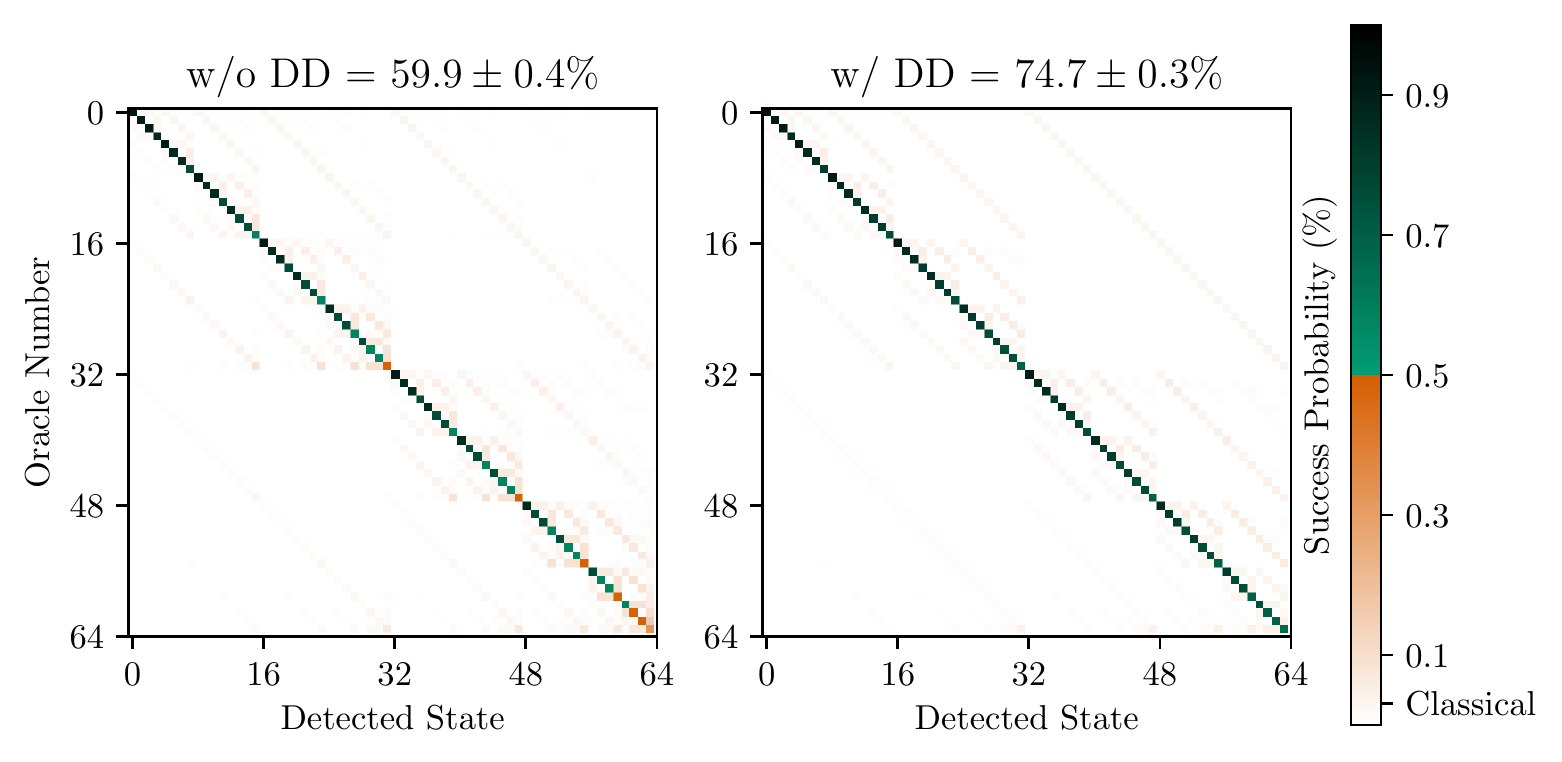} 
\caption{Full output distribution for BV-$6$ from Cairo. Oracles $f_b$ are numbered from $0$ to $63$, corresponding to $b\in \{0^6,\dots,1^6\}$. Ideally, the output state for oracle $f_b$ (vertical axis) is $b$, but in reality, other bitstrings (horizontal axis) are observed as well. Green dots on the diagonal correspond to $p_s>1/2$, where $p_s$ is the empirical frequency (success probability) with which $b$ was output for oracle $f_b$. Without DD, the average success probability over all oracles is $59.9 \pm 0.4\%$, but for some oracles, $p_s<1/2$. With DD, the average success probability rises to $74.7\pm 0.3\%$ and $p_s>1/2$ for all oracles. Success probabilities are reported with $5\sigma$ confidence intervals. }
\label{fig:bv}
\end{figure}

\noindent\textbf{Experimental implementation}\\
\noindent We implemented ssBV-$n$ on two different $27$-qubit QCs: ibmq\_montreal and ibmq\_cairo
%~\cite{IBMQuantum2022} 
(henceforth Montreal and Cairo). Considering that one qubit must function as the ancilla in the BV algorithm, we have $n_{\max}=26$. These devices feature fixed-frequency transmon qubits with microwave pulses for implementing control and readout. While similar in their connectivity, they have different quantum volumes, qubit generations, and gate fidelities (see \cref{app:C} for details).

Setting aside connectivity considerations, given the unknown string $b$ the BV oracle is implemented by performing CNOTs from a subset of the first $|b|=n$ qubits to the ancilla qubit (always numbered $n+1$), preceded and followed by Hadamard gates (see~\cref{fig:bv_circuit}). 
Whether a CNOT is applied from qubit $i$ to the ancilla is determined by whether $b_i=0$ (do nothing) or $b_i=1$ (apply a CNOT). The number of CNOTs is thus the Hamming weight $k$ (the number of $1$'s in $b$). 
Given $n$ and $k$, there are $\binom{n}{k}$ distinct bitstrings, and the circuits for all these bitstrings are identical up to qubit permutation. We exploited this symmetry and generally avoided the impractical task of testing $\sum_{k=0}^n\binom{n}{k}=2^n$ oracles by testing only the $n+1$ cases $b=1^k0^{n-k}$ with $0\le k \le n$ for each $n$.  \\

\noindent\textbf{Results}\\
\noindent BV-$6$ test results, both with and without DD, are shown in~\cref{fig:bv}.  With DD, for every one of the $2^6$ oracles (inputs) tested, the single-shot output success probability exceeds $1/2$, which allows reaching the $2/3$ bounded-error quantum polynomial (BQP) threshold for all possible inputs by classical majority vote on multiple repetitions~\cite{bernsteinQuantumComplexityTheory1997,wrightBenchmarking11qubitQuantum2019}. Without DD, the single-shot output success probability is below $1/2$ for nearly a third of all inputs (though always higher than the classical single-shot probability of $1/64$), so the BQP threshold cannot be reached. This already suggests that error suppression through DD will be central to our quantum speedup demonstration. However, to genuinely demonstrate a quantum speedup we must demonstrate a scaling advantage with problem size $n$, which brings us to our central result.

\begin{figure}[t]
\centering

\includegraphics[trim = 1 0 0 0, clip, width=\columnwidth, valign=t]{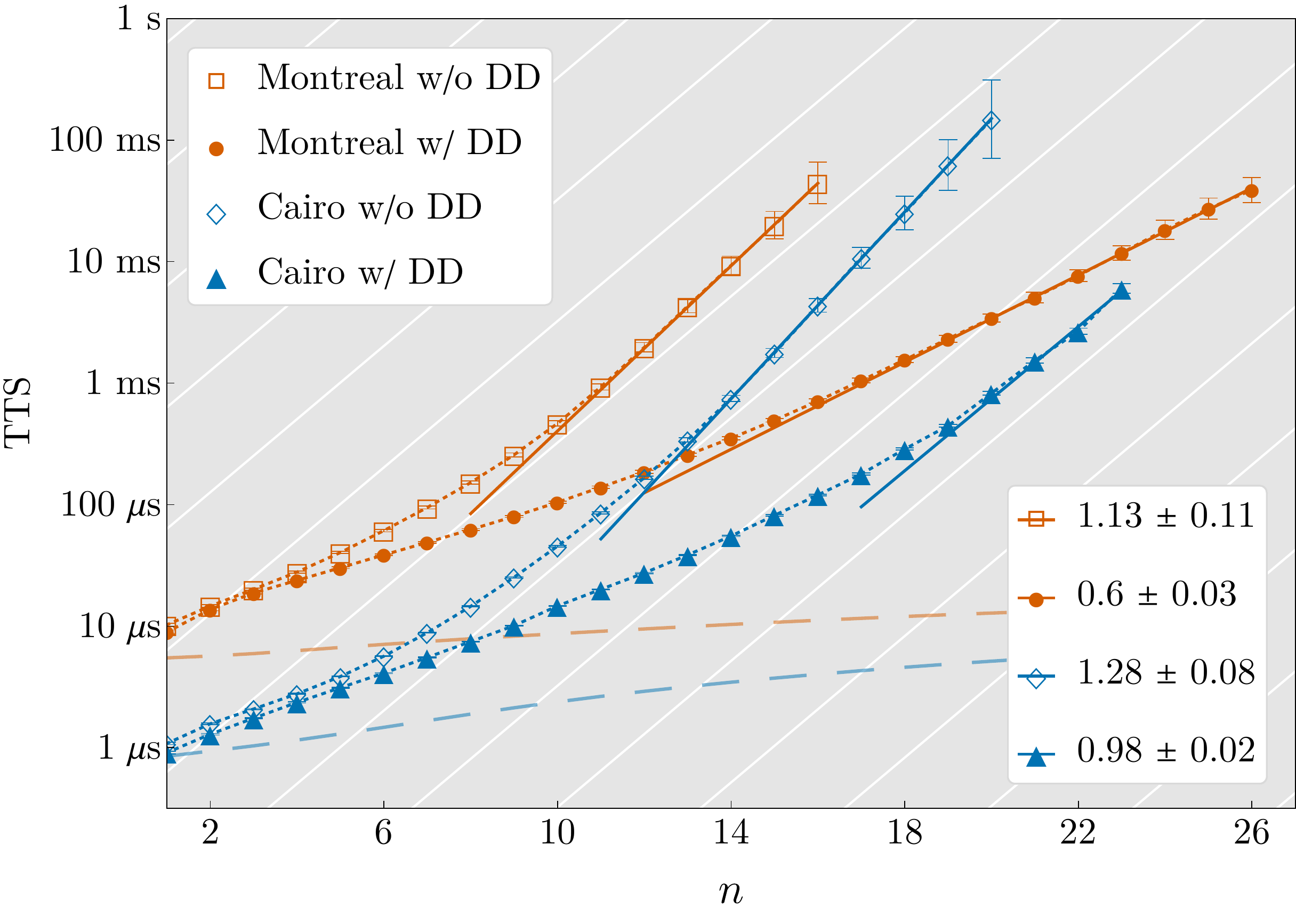} 

\caption{Time-to-solution (TTS) as a function of problem size or number of data qubits $n$. Results for Montreal and Cairo are shown by the orange and blue symbols, respectively, and filled (empty) symbols represent results with (without) DD; dotted lines are guides to the eye. The asymptotic classical scaling $\text{TTS}_C(n)\sim 2^n$ is shown as white grid lines, and the hypothetical, ideal quantum scaling $\text{TTS}_Q(n) \propto n $ of each QC is indicated by the dashed lines (for QC-specific parameter values see \cref{app:C}).
The worst-case scaling fit for each curve is shown by the solid lines, whose slopes $\lambda$ are reported in the bottom legend, with uncertainties representing $95\%$ confidence intervals. Without DD, the TTS results terminate at $n'_{\max}=16$ ($n'_{\max}=20$) for Montreal (Cairo) as $p_s=0$ for $n>n'_{\max}$. Moreover, $\lambda >1$ without DD, indicating a worse-than-classical scaling. With DD protection, on Cairo, the $p_s>0$ range is extended to $n=23$, and $\lambda$ is just below the breakeven point of $1$, but the uncertainty is too large to conclude that quantum speedup has occurred. In contrast, the Montreal scaling with DD does exhibit quantum speedup, as explained in the text. Since two-qubit operations and readout durations are shorter for Cairo, it exhibits a consistently lower absolute TTS than Montreal.  We report $5\sigma$ confidence intervals from bootstrapping for each data point (see Methods); error bars are mostly covered by the symbols.}
 \label{fig:tts}
\end{figure}

Our main result is presented in~\cref{fig:tts}, which shows the TTS as a function of problem size $n$ for both Montreal and Cairo. White grid lines show the asymptotic classical TTS [scaling as $O(2^{n})$], and the ideal quantum TTS (equal to $t_r(n) \sim n$) is shown for reference by the two dashed lines -- one each for Montreal and Cairo. As is apparent visually, the scaling without DD (empty symbols) for both devices is worse than the classical scaling. Moreover, without DD, the success probability drops to zero for $n>16$ (Montreal) and $n>20$ (Cairo); hence the TTS curves terminate. We attribute this worse-than-classical scaling at large problem sizes to the fact that transmon-based devices suffer from spontaneous emission errors, as a result of which they  preferentially generate bitstrings with low Hamming weight, which is worse than a uniformly random guess. This is also apparent from~\cref{fig:bv}(left), which exhibits a trend of lower success probabilities for oracles with a higher index and hence higher Hamming weight.

With DD, this problem is mitigated, so that $p_s>0$ is extended for Cairo (blue) to $n=23$ (excessive readout noise required us to treat Cairo as a device with $n_{\text{tot}}=24$; see \cref{app:C}).
Most notably, it is clear that with DD the Montreal scaling (orange) is better than classical and extends to $n=n_{\max}=26$, suggesting a quantum speedup. 

To quantify this and extract the speedup exponent $\lambda$ as conservatively as possible, we compute the worst-case scaling from our data (see Methods). 
The results are shown as the straight blue and orange lines in~\cref{fig:tts}, along with the numerical values of $\lambda$ in the legend. Without DD, we obtain $\lambda=1.13 \pm 0.11$ and $1.28 \pm 0.08$ for Montreal and Cairo, respectively, meaning a quantum slowdown. For Cairo, the scaling with DD is $\lambda = 0.98 \pm 0.02$, not a statistically significant difference from the classical scaling. However, the fit confirms that \emph{Montreal with DD exhibits a quantum speedup}: $\lambda=0.60 \pm 0.03$. All the reported uncertainties represent $2\sigma$ symmetric confidence intervals; see Methods. The difference between Cairo and Montreal agrees with the reported larger quantum volume ($128$ \textit{vs} $64$) of Montreal~\cite{Pelofske:2022}, and suggests that the latter is a relevant performance metric also in the present context of algorithmic speedups.

\begin{figure}[t]
\centering
\includegraphics[trim = 0 0 0 0, clip, width=\columnwidth, valign=t]{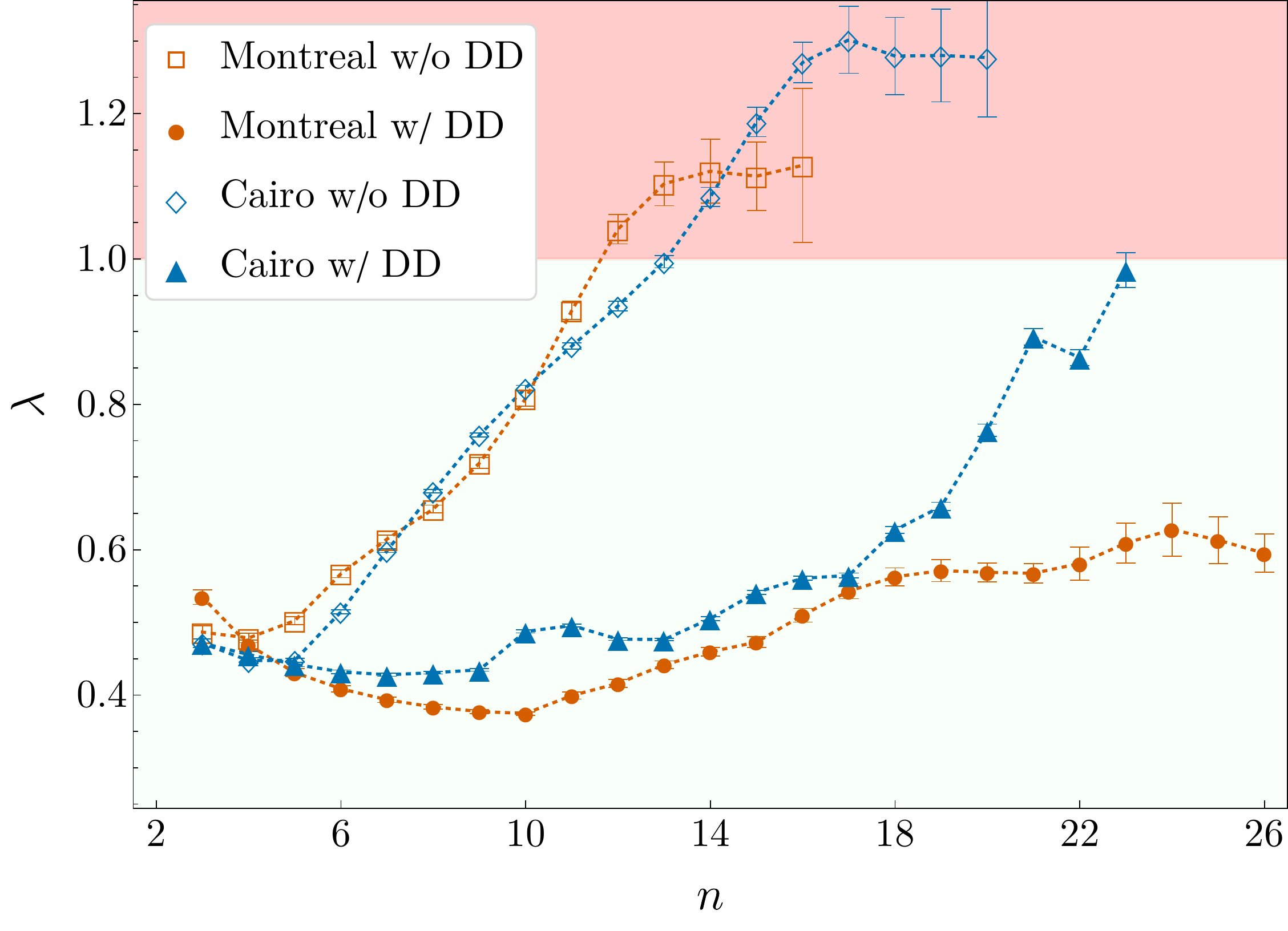}
\caption{ Results for $\lambda_{h_{\max}}$, the maximum local slope of each of the curves in~\cref{fig:tts} for $n\leq h_{\max}$, i.e., the worst-case-scaling when restricted to QC with $h_{\max}$ + 1 qubits. Only Montreal with DD exhibits an unambiguous quantum speedup, with $\lambda_{h_{\max}}$ well below $1$ for all $n\leq h_{\max}$. Error bars represent $2\sigma$ confidence intervals.}
 \label{fig:tts2}
\end{figure}

It is clear from~\cref{fig:tts} that all the slopes vary with $n$. One might thus ask what the scaling would appear to be for a hypothetical QC with fewer qubits ($h_{\max}$) than the actual $n_{\max}=26$; we address this in~\cref{fig:tts2}.  This figure shows the maximum local slope of each of the curves in~\cref{fig:tts} for $n\leq h_{\max}$ (see Methods). The results clearly show the growth of the speedup exponents $\lambda_{h_{\max}}$ for Cairo with and without DD, and for Montreal without DD, to the point $\lambda > 1$ or beyond, where no quantum speedup survives. In contrast, the speedup exponent for Montreal with DD is well within the quantum speedup region of $\lambda <1$ for all values of $h_{\max}$. \\

\noindent\textbf{Discussion and Conclusions}\\
\noindent The ssBV-$n$ problem has a \emph{provable} exponential quantum speedup over the best possible classical algorithm in the setting of a game involving an oracle and a verifier, which makes it an attractive target for an experimental quantum speedup demonstration.
The BV algorithm, the key ingredient in our work, is frequently used for benchmarking NISQ devices~\cite{wrightBenchmarking11qubitQuantum2019,  royProgrammableSuperconductingProcessor2020, debnathDemonstrationSmallProgrammable2016, linkeExperimentalComparisonTwo2017, muraliFullstackRealsystemQuantum2019}, with the largest prior implementations being BV-$11$ and BV-$20$~\cite{lubinskiApplicationOrientedPerformanceBenchmarks2021} on superconducting and trapped-ion devices respectively. Here we reached BV-$26$ on a superconducting device (ibmq\_montreal), by invoking dynamical decoupling (DD). This corresponds to a maximal circuit depth of $44$ CNOTs ($46$ when accounting for the Hadamard gates at the beginning and end), slightly higher than the largest depth ($40$) reached in quantum supremacy experiments~\cite{Arute:2019aa,wu2021strong}; see \cref{app:D}.

To test for a quantum speedup, we compared the asymptotic scaling of the TTS metric with problem size for both classical and quantum algorithms. We demonstrated a statistically significant algorithmic quantum speedup on Montreal using this metric. A crucial feature in our implementation was error-suppression through DD, without which the speedup was not exhibited. 

It is reasonable to question whether this speedup can be expected to continue indefinitely. In the absence of fault-tolerant quantum error correction~\cite{Campbell:2017aa} decoherence always eventually dominates, and one should expect the DD-enabled quantum speedup to have a finite-sized upper limit. The fact that this upper limit is not observed in our experiments satisfies a key goal of implementing a quantum algorithm on a NISQ device, namely to check whether a quantum advantage is already accessible even before the advent of fault-tolerance, up to the largest problem sizes supported by the device. We have shown here that, with the help of error suppression via DD, this is indeed the case. 

Another natural question is to what extent the speedup reported here can be further improved. We certainly expect that methods such as error mitigation~\cite{Temme:2017aa} and further DD sequence optimization~\cite{alex2020deep,raviVAQEMVariationalApproach2021, dasADAPTMitigatingIdling2021,niu2022analyzing} will have such an effect, though TTS$_Q$ should then account for the additional classical computation time they incur. Device-tailored optimization of DD sequences with advanced low-level pulse control is an exciting frontier that remains largely unexplored and appears particularly promising. While we focused on superconducting-qubit devices, DD-protection can be beneficial across platforms, as all NISQ devices are affected by computational errors such as decoherence and crosstalk. 

An ideal quantum computer would reduce the exponential growth of the classical TTS for the ssBV-$n$ problem to a linear one. Our results are comparatively less impressive: we demonstrated what amounts to a polynomial quantum speedup, by reducing the exponent of the TTS scaling below its classical minimum. Our work provides a path to testing such speedups across platforms and algorithms in the NISQ era.\\\\

\noindent\textbf{Methods}\\

\begin{figure}[h]
\centering
\begin{tabular}{cc}

\Qcircuit @C=1em @R=0.9em @!R {
\lstick{| 0 \rangle}  & \gate{H} & \ctrl{4} & \qw   & \gate{H} & \meter   \\
\lstick{| 0 \rangle}  & \gate{H} & \qw & \ctrl{3}  & \gate{H} & \meter  \\
\lstick{| 0 \rangle}  & \qw & \qw & \qw & \qw & \qw  \\
\lstick{| 0 \rangle}  & \qw & \qw & \qw   & \qw & \qw  \\
\lstick{| 1 \rangle}  & \gate{H}  & \targ & \targ & \gate{H}  & \meter
}

&
\hspace{6mm}

\Qcircuit @C=1em @R=0.9em @!R {
\lstick{| 0 \rangle}  & \gate{H} & \ctrl{4} & \qw   & \gate{H} & \meter   \\
\lstick{| 0 \rangle}  & \gate{H} & \qw & \ctrl{3}  & \gate{H} & \meter  \\
\lstick{| 0 \rangle}  & \gate{H} & \qw & \qw & \gate{H} & \meter   \\
\lstick{| 0 \rangle}  & \gate{H} & \qw & \qw   & \gate{H} & \meter  \\
\lstick{| 1 \rangle}  & \gate{H}  & \targ & \targ & \gate{H}  & \meter
}

\end{tabular}
\caption{Equivalent circuits used in our reduction from the circuit for $b=1^k0^{n-k}$ to the circuits for $b=1^k0^{m-k}$ with $m\in[k,n-1]$. Illustrated on the left is BV-$2$ with the $b=11$ oracle ($m=2$ and $k=2$), and on the right BV-$4$ with the $b=1100$ oracle ($n=4$ and $k=2$). The left circuit is obtained from the right circuit by tracing over the last $n-k=2$ data qubits. }
\label{fig:equiv}
\end{figure}
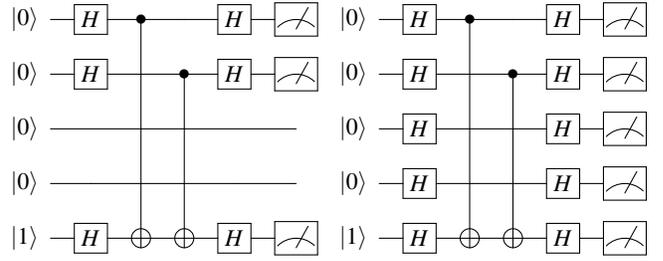

\noindent \textsc{Reduction from ssBV-$n$ to ssBV-$m$}\\
\noindent Since the oracle acts trivially on the $i$-th qubit if $b_i = 0$, the Hadamard gate pairs on the last $n-k$ qubits cancel.  
Therefore, the only difference between $b=1^k0^{n-k}$ and $b=1^k0^{m-k}$ is that the BV circuit for $|b|=n$ or $|b|=m$ applies \emph{cancelling} Hadamard gates to the last $n-k$ or $m-k$ qubits, respectively. Now let $m\in [k,n-1]$; then all the circuits for $b=1^k0^{m-k}$ in theory have the identical output as the circuit for $b=1^k0^{n-k}$, as illustrated in Fig.~\ref{fig:equiv}.
We may thus extract the ssBV-$m$ results from the ssBV-$n$ results by running only the BV-$n$ circuits and tracing over the last $n-m$ data qubits (see \cref{app:B} for a detailed proof of this statement, including in the open system case), a practice we implemented in our experiments and subsequent analysis.
\\

\noindent \textsc{Circuit depth reduction}\\
\noindent A crucial ingredient in our implementation is a simple method for circuit depth reduction, needed to overcome the limited connectivity of the Montreal and Cairo chip architectures, where most qubits have only two neighbors, and some have three. This necessitates many SWAP operations in order to implement a CNOT between non-neighboring qubits, as required according to~\cref{fig:bv_circuit}. While implementing a SWAP requires three CNOTs, by using the circuit identity $\text{CNOT}_{12} \  \text{SWAP}_{12} = \text{CNOT}_{21} \text{CNOT}_{12}$, we can implement a CNOT followed by a SWAP with just two CNOTs. On the heavy-hex layout of Montreal and Cairo, we found that the number of CNOTs required for implementing ssBV-$n$ in this manner scales as $1.76n$. Our longest circuit, ssBV-$26$, required a total of $44$ CNOTs, reduced from 80 CNOTs had we used the standard SWAP method. See \cref{app:D} for more details.  
\\

\noindent \textsc{Data Collection and error analysis}\\
\noindent We implemented each unique circuit for 100,000 shots and 32,000 shots on Cairo and Montreal, respectively. We then sampled the corresponding results for all BV-$n$ oracles using bootstrapping~\cite{efronBootstrapMethodsAnother1992} and report the mean TTS for BV-$n$ along with error bars corresponding to $\pm 5 \sigma$ for the bootstrapped distribution (see \cref{app:E} for more details).
\\

\noindent \textsc{Computation of the Worst-Case TTS Scaling}\\
\noindent To compute the scaling exponents reported in~\cref{fig:tts}, we first fit $\text{TTS} \propto 2^{\lambda_{l,u} n}$ to the data for $n\in [l,u]$ with $l\in [1,u-2]$ and $ u \in [3,n'_{\max}]$, where $n'_{\max}$ is the largest $n$ for which we have TTS data, i.e., $n'_{\max}\in\{16,20,23,26\}$ for $\{$Montreal w/o DD, Cairo w/o DD, Cairo w/DD, Montreal w/DD$\}$, respectively. In other words, $\lambda_{l,n'_{\max}}$ is the slope obtained by fitting to the data between $l$ and $n'_{\max}$.
We then obtain the asymptotic speedup exponent via $\lambda = \max_{l} \ \lambda_{l, n'_{\max}}$. By taking the $\max$, we ensure that the speedup exponent thus estimated is conservative, i.e., represents the worst-case scaling compatible with the data. Similarly, \cref{fig:tts2} shows the speedup exponent $\lambda_{h_{\max}}= \max_{l} \ \lambda_{l,h_{\max}}$, i.e., the maximum local derivative of each of the curves in~\cref{fig:tts} for $n\leq h_{\max}$.  All $\lambda$ values are reported with a $2\sigma$ confidence interval obtained after bootstrapping.  For complete details see \cref{app:E}. \\

\noindent \textsc{DD sequence placement}\\
\noindent The placement of DD sequences is determined by idle times (gaps) in the quantum circuit. These gaps arise because the algorithm specifies them or due to limited connectivity of the underlying architecture, which requires information swapping between some qubits while others are idle. Here we implemented one repetition of a pulse sequence that fits precisely in a gap, illustrated schematically in~\cref{fig:bv_circuit}(bottom). Based on the results of the survey of DD sequences~\cite{DD-survey}, we used the universally robust ($\text{UR}_n$~\cite{genovArbitrarilyAccuratePulse2017}) sequences UR$_{14}$ and UR$_{18}$ on Montreal and Cairo, respectively. See \cref{app:F} for details.

\section*{Acknowledgements}
Access to the IBM Quantum Network was obtained through the membership of UNM in the IBM Quantum Hub at NC State. We are grateful to UNM for generously providing us access through their membership. This material is based in part upon work supported by the National Science Foundation the Quantum Leap Big Idea under Grant No. OMA-1936388. This work is also partially supported by a DOE/HEP QuantISED program grant, QCCFP/Quantum Machine Learning and Quantum Computation Frameworks (QCCFP-QMLQCF) for HEP, Grant No. DE-SC0019219.
We are grateful to Dr. Namit Anand, Dr. Victor Kasatkin, and Dr. Evgeny Mozgunov for useful comments on the manuscript. 

\section*{Data and code availability}

The data supporting the findings of this paper are available at \href{https://tinyurl.com/ssBV-data}{https://tinyurl.com/ssBV-data}.
The code used in this paper is available at \href{https://github.com/USCqserver/bv-quantum-speedup}{github.com/USCqserver/bv-quantum-speedup}.

\appendix 

\section{Prior better-than-classical results}
\label{app:A}

Quantum supremacy~\cite{Preskill:2012aa} has been demonstrated for efficiently sampling pseudo-random quantum circuits~\cite{Arute:2019aa,wu2021strong} and boson sampling~\cite{Zhong:2021wv}. Assuming that it takes exponential time to solve an NP-complete problem, a quantum advantage in an interactive prover-verifier setting was demonstrated in Ref.~\cite{centroneExperimentalDemonstrationQuantum2021}. These results all rely on various complexity-theoretic conjectures~\cite{Aaronson:2016aa,Harrow:2017aa,Lund:2017aa} to perform computational tasks that are beyond the capability of any classical computer. They are also susceptible to the boundaries defined by the capabilities of ever-improving classical supercomputers and algorithms~\cite{Zlokapa:2020}.

Quantum annealing hardware based on superconducting flux qubits was used to demonstrate a speedup against simulated annealing for certain specially crafted spin-glass problems~\cite{Albash:2017aa}. However, the speedup did not hold against other classical algorithms (such as quantum Monte Carlo). Another recent experiment used a variational quantum adiabatic algorithm, implemented using Rydberg atom arrays, to demonstrate a solution of the Maximum Independent Set problem with a quantum advantage as a function of problem hardness (not problem size) against simulated annealing~\cite{Ebadi:22}. The latter was constrained to the same effective circuit depth as the quantum implementation. The speedup was not tested against state-of-the-art classical algorithms, such as parallel tempering with iso-energetic cluster moves~\cite{PhysRevLett.115.077201}. 

A quantum linear optics device solved the quantum coupon collection problem in fewer attempts than a classical strategy~\cite{zhouExperimentalQuantumAdvantage2022}. Likewise, using a nanophotonic device, a quantum speedup in learning time was demonstrated in a reinforcement learning setup~\cite{saggioExperimentalQuantumSpeedup2021}. However, these demonstrations do not scale: in both cases the authors show that it is optimal to revert to the classical strategy even for the cases they tested, when the problem sizes or learning times increase beyond a certain threshold. 

Recent work used hybrid quantum-classical learning with quantum-enhanced experiments to demonstrate an experimental quantum advantage~\cite{Huang:2021}. This works proves an exponential quantum-classical separation in terms of the number of experiments required to achieve a given accuracy in the machine learning task, and demonstrates this separation experimentally. In a similar vein, a quantum advantage was demonstrated in a setting of supervised learning assisted by an entangled sensor network against classical support vector machines (an entanglement-enabled reduction in the error probability for classification of
multidimensional radio-frequency signals)~\cite{xiaQuantumEnhancedDataClassification2021}.
The most important difference is that in both cases this involving quantum data (i.e., learning on quantum states), whereas in our case the problem being solved is classical: winning a guessing game involving purely classical data (secret bitstrings). Another important difference between these results and ours is that the former are hybrid quantum-classical approaches, whereas we are considering the scaling of a purely quantum algorithm. The difference is significant in that in the former case the cost of the classical processing (such as variational optimization) is not fully accounted for. We do not expect this cost to change an exponential separation, but this point makes a direct comparison with our result slightly more challenging. 

Some of our results are directly comparable to various NISQ implementations of oracular quantum algorithms. Ref.~\cite{wrightBenchmarking11qubitQuantum2019} implemented the BV and Hidden Shift (HS) algorithms for $n=10$ on an 11-qubit trapped-ion device, which was the largest and most successful implementation of these algorithms at the time (later superseded by BV-$11$ and BV-$20$ on superconducting and trapped-ion devices respectively~\cite{lubinskiApplicationOrientedPerformanceBenchmarks2021}). Moreover, they demonstrated better-than-classical performance on both algorithms, by crossing the BQP threshold for BV  and by finding higher-than-classical success probabilities for HS, at the fixed problem size of $n=10$. Likewise, Refs~\cite{figgattComplete3QubitGrover2017} and~\cite{royProgrammableSuperconductingProcessor2020} respectively used 3-qubit trapped-ion and trimon devices to implement oracular algorithms, and 4-qubit and 5-qubit versions of Grover's algorithm were implemented on various IBMQ devices~\cite{Zhang_2021}. In all these cases claims of better-than-classical results focus on crossing a classical success probability threshold at a fixed problem size such that $p_{\text{quantum}} > p_{\text{classical}}$. However, this argument is insufficient for establishing a quantum speedup, which must be based on the scaling with $n$ of a time-based metric such as the TTS~\cite{ronnowDefiningDetectingQuantum2014} or related metrics~\cite{King:2015cs,Vinci:2016tg,2016arXiv160401746M}.

\section{Reduction from ssBV-\texorpdfstring{$n$}{n} to ssBV-\texorpdfstring{$n$}{m} for CPTP maps}
\label{app:B}

We extracted the ssBV-$m$ results  (for $m<n$) from the ssBV-$n$ results by tracing over the last $n-m$ qubits. Here we prove the equivalence of our procedure to actually running the ssBV-$m$ circuits, as long as the completely positive, trace preserving (CPTP) map governing the circuit in the open system case factors into a product over the ``marked'' and ``unmarked'' qubits, i.e., those corresponding to a $1$ (marked) or $0$ (unmarked) in the bitstring $b$ that defines the given oracle. 

We first consider the ideal ssBV-n algorithm without any open-system effects. The initial state is $\left|\psi_{0}\right\rangle=|0\rangle^{\otimes n} \otimes |1\rangle$.
Applying the initial Hadamard layer yields
\beq
|\psi_{1}\rangle=(H^{\otimes n+1})|\psi_{0} \rangle=|+\rangle^{\otimes n} \otimes |-\rangle .
\eeq
We then apply the oracle corresponding to the hidden bitstring $b\in\{0,1\}^n$, i.e., $O_{f_b} | x \rangle |y \rangle=| x \rangle | y \oplus f_b(x) \rangle$, so that the state becomes
\bes
\begin{align} 
| \psi_{2} \rangle  &= O_{f_b}  |\psi_{1} \rangle =
\\
&=O_{f_b}\left[\frac{1}{\sqrt{2^{n+1}}} \sum_{x \in\{0,1\}^{n}}|x\rangle(|0\rangle-|1\rangle)\right]  \\
& =\frac{1}{\sqrt{2^{n+1}}} \sum_{x \in\{0,1\}^{n}} |x\rangle(|f_b(x)\rangle-|1 \oplus f_b(x)\rangle) \\
& = \frac{1}{\sqrt{2^{n}}} \sum_{x}(-1)^{f_b(x)}|x\rangle | - \rangle .
\end{align}
\ees
In the last line above, we use the fact that for any $x$, $f_b(x)$ is either 0 or 1, i.e., 
\bes
\begin{align}
|x\rangle(|f_b(x)-|1 \oplus f_b(x)\rangle) & =	\left\{	\begin{array}{ll} | x \rangle (|0|-|1\rangle) & f_b(x)=0 \\ 	| x \rangle(|1\rangle-|0\rangle) & f_b(x)=1 	\end{array}\right. \\ 
& =(-1)^{f_b(x)}|x\rangle(|0\rangle-|1\rangle) .
\end{align}
\ees
The final Hadamard layer is applied next, and, using $f_b(x) = b \cdot x$ (mod 2):
\bes
\label{eq:psi3}
\begin{align} 
| \psi_{3} \rangle  &= H^{\otimes n+1} |\psi_{2} \rangle \\
& =\left(H^{\otimes n} \frac{1}{\sqrt{2^{n}}} \sum_{x}(-1)^{b \cdot x} | x \rangle \right) \otimes  H | - \rangle =|b\rangle|1\rangle .
\end{align} 
\ees

In preparation of our more general discussion below, let us equivalently represent the action of ssBV-$n$ with hidden bitstring $b$ on some initial state $\rho$ of the $n$ data qubits and the ancilla qubit as
\beq
\BV_n(b)[\rho] = \text{Tr}_{n+1}\left[ \left( H^{\otimes n+1} \circ O_{f_b} \circ H^{\otimes n+1} \right) [\rho] \right] ,
\eeq
where $O_{f_b}$ represents the BV oracle, and $\text{Tr}_{n+1}$ means that the state of the ancilla qubit (numbered $n+1$) is discarded at the end, so that $\BV_n(b)[\rho]$ is the state of the $n$ data qubits at the end of one run of the algorithm. Thus, if we write $\Pi_b = | b \rangle\! \langle b |$ and $| \psi_0 \rangle\! \langle \psi_0 | = \Pi_{0^n} \otimes \Pi_1$, then it follows from~\cref{eq:psi3} that $\BV_n(b)[| \psi_0 \rangle\! \langle \psi_0 |] =  \Pi_b$. In the main text and Methods we argued that if $b_n = 1^k0^{n-k}$, then 
\beq
\BV_m(b_m)[\Pi_{0^m} \otimes \Pi_1] = \text{Tr}_{\{ m+1, m+2, \dots, n \}} \left[  \BV_n(b_n)[\Pi_{0^n} \otimes \Pi_1] \right] ,
\label{eq:BV-equiv}
\eeq
where the trace means that the states of qubits $\{ m+1, m+2, \dots, n \}$ are discarded from the result of running $\BV_n(b_n)$. To prove this claim in the absence of any noise, note that:
\bes
\begin{align} 
\text{Tr}_{\{ m+1, m+2 \dots n \}} & \left[  \BV_n(b_n)[| \psi_0 \rangle\! \langle \psi_0 |] \right] \\
&=\text{Tr}_{\{ m+1, m+2 \dots n \}} \left[ \Pi_{b_{n}} \right]\\
&=\text{Tr}_{\{ m+1, m+2 \dots n \}} \left[ \Pi_{1^k0^{n-k}} \right]\\
&=\text{Tr}_{\{ m+1, m+2 \dots n \}} \left[ \Pi_{1^k0^{m-k}} \otimes \Pi_{0^{n-m}}   \right]\\
&= \Pi_{1^k0^{m-k}}  \\
&= \BV_m(b_m)[\Pi_{0^m} \otimes \Pi_1] ,
\end{align} 
\ees
as claimed.

Now consider the case where each gate is represented not by a unitary but a CPTP map. For convenience, let us renumber the ancilla to be the $0$'th qubit instead of the $n+1$'th qubit. We can rewrite the initial state as
\beq 
| \psi_0 \rangle\! \langle \psi_0 |  = \Pi_1 \otimes \Pi_{0^n}  =  \rho_A \otimes \rho_B ,
\eeq
where $\rho_A=(\Pi_1 \otimes \Pi_{0^m})$ and $\rho_B = \Pi_{0^{n-m}}$.
The BV oracle does not introduce any two-qubit gates between qubit sectors $A=\{ 0,1,..m \}$ and $B=\{ m+1,m+2,..n \}$, so it is reasonable to assume that under the coupling to the environment they remain uncoupled (as long as there is no unintended crosstalk between the two sectors). Therefore, the CPTP map for the noisy ssBV-$n$ algorithm will be
\begin{align}
\mathcal{BV}_n(b_n)[\rho_A \otimes \rho_B]  &= \text{Tr}_{0}\left[ (\mathcal{H}_A \otimes \mathcal{H}_B) \circ  (\mathcal{O}_A \otimes \mathcal{O}_B) \right. \notag \\
&\qquad\qquad \left. \circ  (\mathcal{H}_A \otimes \mathcal{H}_B)[\rho_A \otimes \rho_B] \right], 
\end{align}
where $\mathcal{H}_{A,B}$ and $\mathcal{O}_{A,B}$ represent the CPTP maps corresponding to the experimental implementation of the unitaries $H$ (multi-qubit Hadamard) and $O$ (oracle) acting on qubit sectors $A,B$. Recall that for arbitrary CPTP maps $\mathcal{U}$ and $\mathcal{V}$ acting on a tensor-product space 
\beq
\mathcal{U} \otimes \mathcal{V}\ [\rho \otimes \sigma] = \mathcal{U}[\rho] \otimes \mathcal{V}[\sigma]. 
\eeq

\begin{figure}[t]
    \centering
    \includegraphics[trim = 0 0 0 0, clip, width=\columnwidth, valign=t]{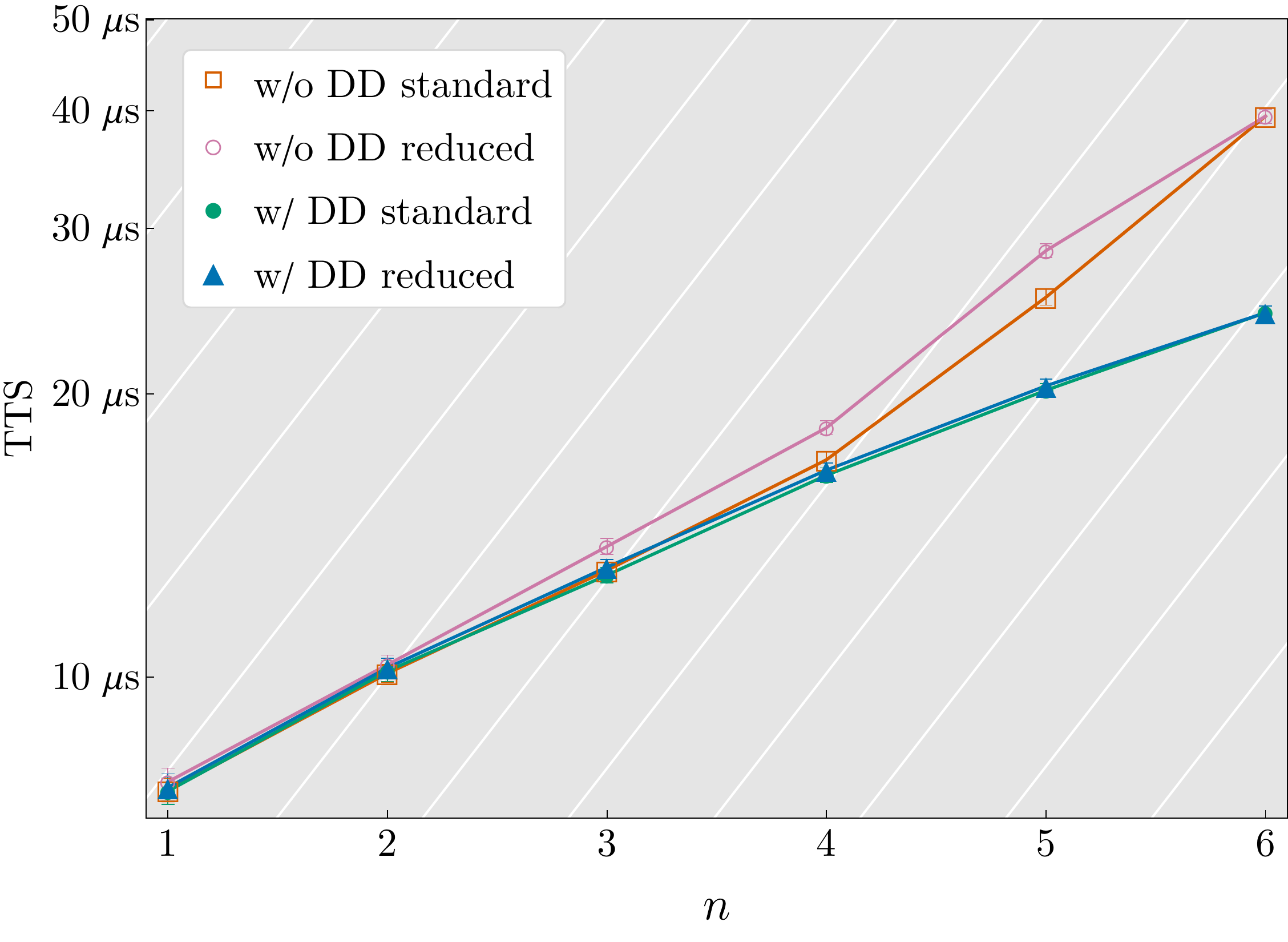} 
    \caption{Comparison of the effect of crosstalk without and with DD (the UR$_{14}$ sequence). $\text{TTS}(n)$ is shown for ibmq\_jakarta. While $\text{TTS}_{\text{standard}} \leq \text{TTS}_{\text{reduced}}$ without DD protection, the TTS results are statistically indistinguishable in the presence of DD.}
     \label{fig:jakarta}
    \end{figure}

\begin{figure}[t]
    \centering
    \includegraphics[width=\columnwidth]{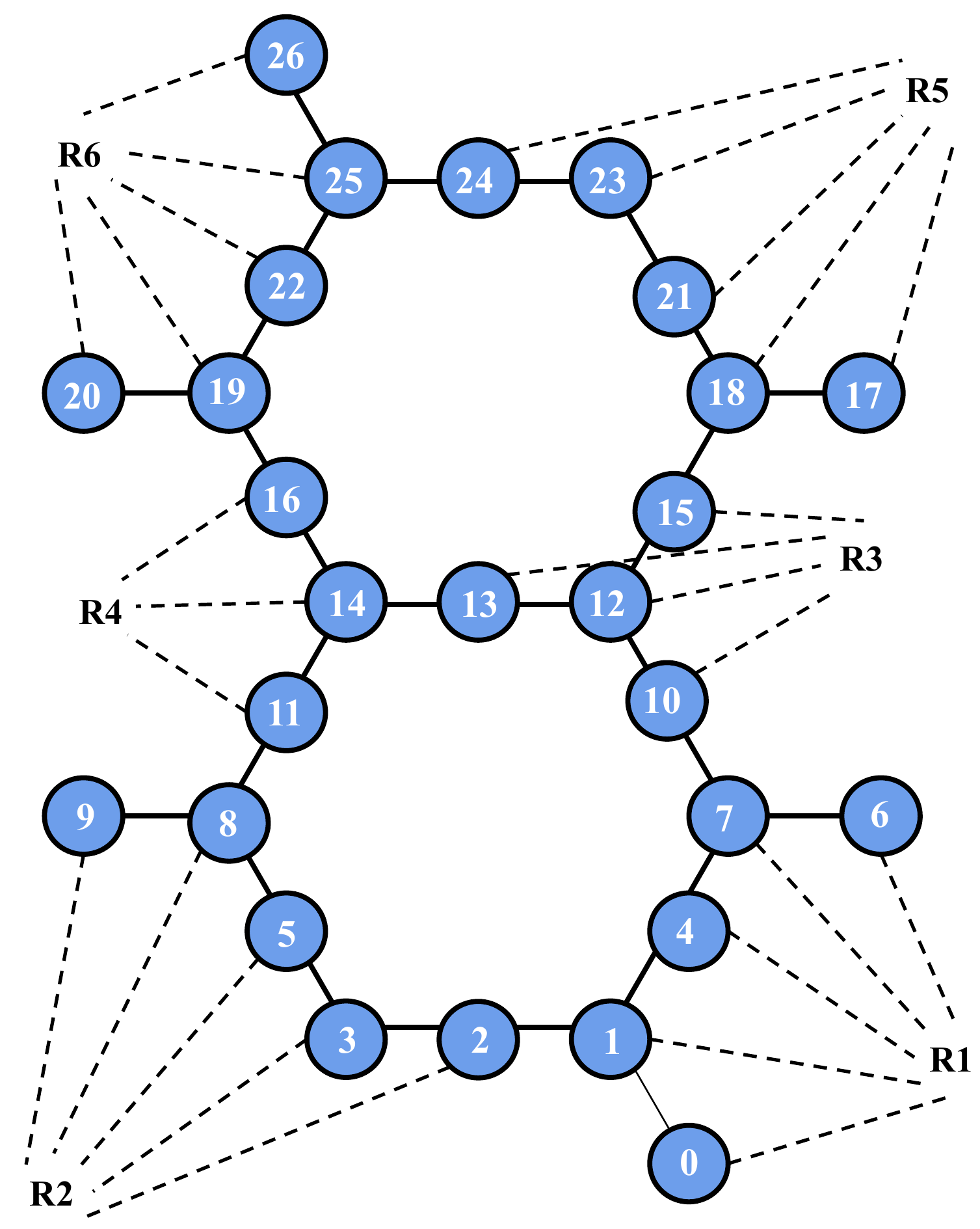} 
    \caption{Schematic of the lattice connectivity for 27-qubit devices with the heavy-hex layout~\cite{jurcevicDemonstrationQuantumVolume2021}. The dashed lines connect qubits that are multiplexed together for readout.}
    \label{fig:architecture}
\end{figure}

\begin{figure}[t]
    \centering
    \includegraphics[trim = 0 0 0 0, clip, width=\columnwidth, valign=t]{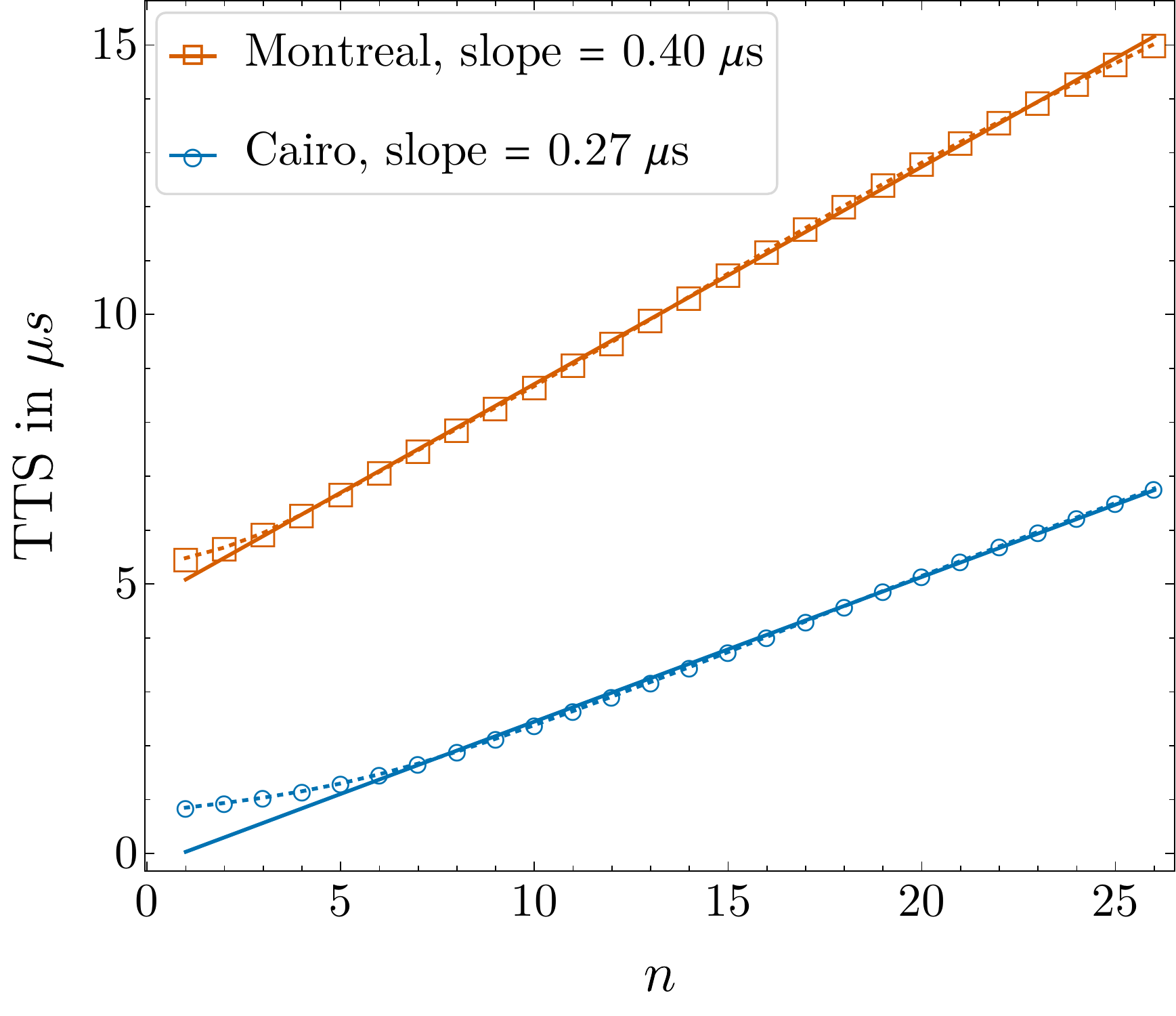} 
    \caption{The hypothetical, ideal TTS for Montreal and Cairo as a function of $n$. A deviation from linear scaling is seen for small $n$. }
     \label{fig:durations_fit}
\end{figure}

\begin{figure}[t]
\centering
\includegraphics[trim = 0 0 0 0, clip, width=\columnwidth, valign=t]{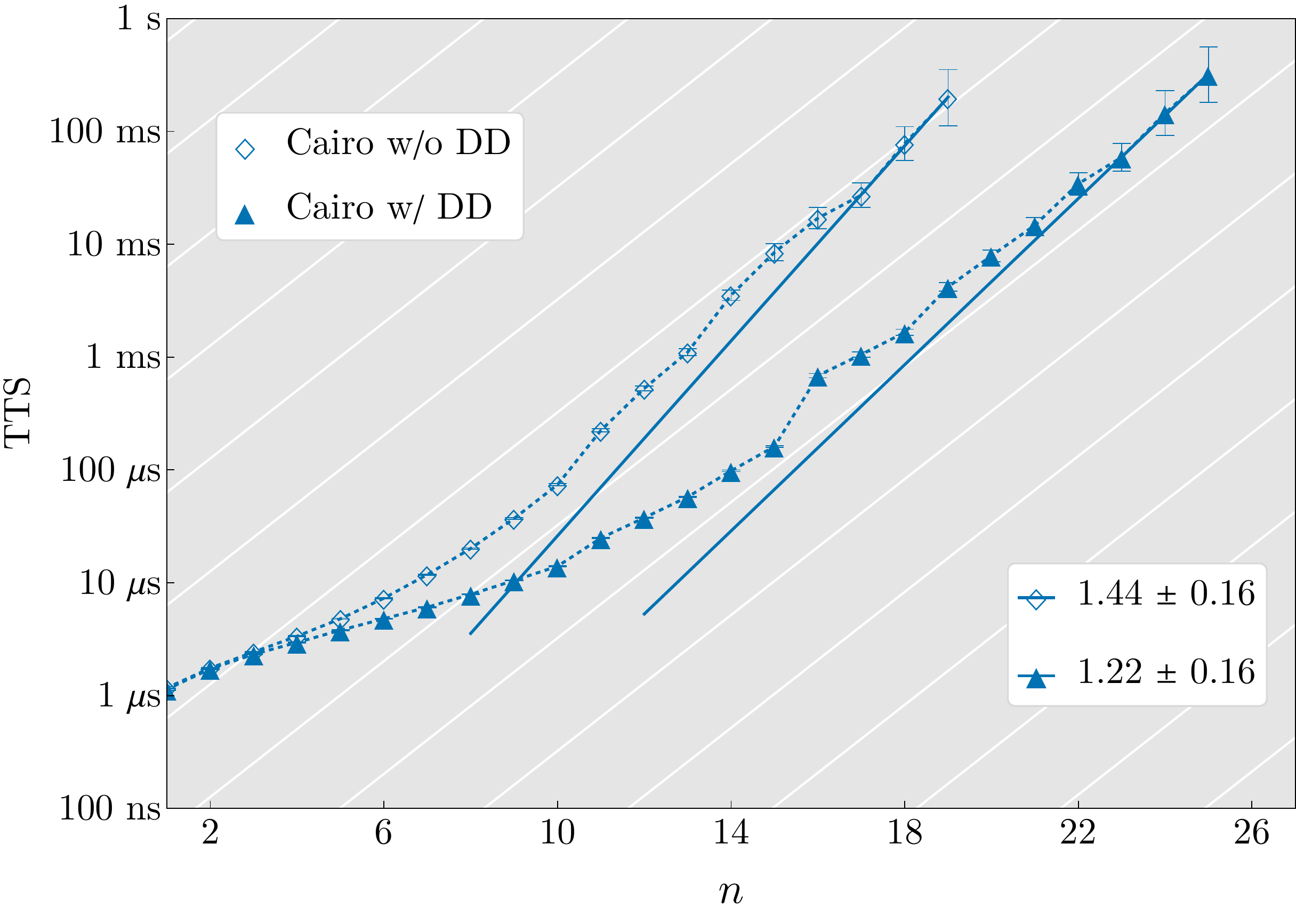} 
\caption{Results from Cairo with a different logical to physical qubit mapping than used in \cref{fig:tts}, using all $27$ qubits. These runs were performed on January 2nd, 2022. Due to a significant readout error in Q19, there is an abrupt jump in the TTS at $n=15$. Consequently, qubits Q19, Q20, and Q22 were left out of all subsequent experiments. }
 \label{fig:cairo_faulty}
\end{figure}

Therefore, 
\bes
\begin{align} 
&\text{Tr}_{\{ B \}} \left[  \mathcal{BV}_n(b_n)[ | \psi_0 \rangle\! \langle \psi_0 | ] \right] \\
&\quad = \text{Tr}_{\{ 0, B \}} \left[  (\mathcal{H}_A \otimes \mathcal{H}_B) \circ  (\mathcal{O}_A \otimes \mathcal{O}_B) \right. \notag \\
&\left. \qquad \qquad\qquad \circ  (\mathcal{H}_A \otimes \mathcal{H}_B) [\rho_A \otimes \rho_B] \right] \\
&\quad = \text{Tr}_{\{ 0, B \}} \left[  (\mathcal{H}_A \circ \mathcal{O}_A \circ \mathcal{H}_A) [\rho_A]\notag \right. \\
&\left. \qquad\qquad\qquad \otimes (\mathcal{H}_B \circ \mathcal{O}_B \circ \mathcal{H}_B) [ \rho_B] \right] \\
&\quad =  \text{Tr}_{\{ 0 \}} \left[  (\mathcal{H}_A \circ \mathcal{O}_A \circ \mathcal{H}_A) [\rho_A] \right]\\
&\quad =  \mathcal{BV}_m(b_m)[\rho_A] \\
&\quad = \mathcal{BV}_m(b_m)[\Pi_1 \otimes \Pi_{0^m}]. 
\end{align}
\ees
This is the CPTP map generalization of the closed-system result~\cref{eq:BV-equiv}, and it shows that that the reduction from ssBV-$n$ to ssBV-$m$ holds rigorously also in the open system setting, as long as as the CPTP map factors according to the qubit sectors $A$ and $B$.

Ref.~\cite{tripathiSuppressionCrosstalkSuperconducting2022} considered in detail how the state of the spectator qubits can change the effect of crosstalk on neighboring qubits. In particular, it showed that while crosstalk can be exacerbated if the spectator qubits are in a superposition state (as is the case here), this effect can be counteracted by using DD. The improvement we have reported under DD further confirms this observation in the context of ssBV-$n$ circuits. 

To investigate this more closely, we  implemented ssBV-$n$ for $n=1$ to $6$ on the 7-qubit ibmq\_jakarta processor, and considered both the standard and the reduced setup (i.e., both sides of~\cref{fig:equiv}. The TTS results are shown in~\cref{fig:jakarta}. Without DD, we expect cross-talk to introduce unwanted coupling between the marked and unmarked qubits, which lowers the success probability and hence increases the TTS. This is what we already observed on Montreal and Cairo, and is also seen on Jakarta. Furthermore, we expect the proof of the reduction from ssBV-$n$ to ssBV-$m$ to break down in the presence of cross-talk, and indeed, we observe in~\cref{fig:jakarta} that $\text{TTS}_{\text{standard}} \neq \text{TTS}_{\text{reduced}}$ without DD-protection. However, with DD the TTS is statistically identical for both setups. This further validates our use of the reduced setup.

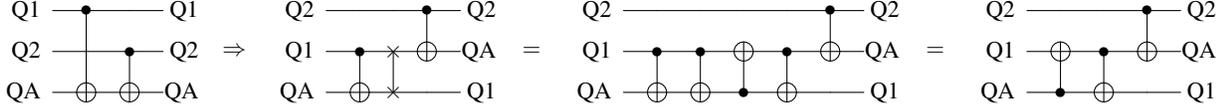
\begin{figure*}[t]
    \centering
    \begin{equation*}
    \begin{tabular}{cccc}
        \centering
        \Qcircuit @C=1em @R=0.9em @!R {
        \lstick{\text{Q1}} & \ctrl{2} & \qw & \qw & & \lstick{\text{Q1}}\\
        \lstick{\text{Q2}} & \qw & \ctrl{1}  & \qw & & \lstick{\text{Q2}} & \Rightarrow \\
        \lstick{\text{QA}} & \targ & \targ  & \qw & & \lstick{\text{QA}}
        } & 
        \hspace{10mm}

        \Qcircuit @C=1em @R=0.9em @!R {
        \lstick{\text{Q2}} & \qw & \qw &\ctrl{1} & \qw & & \lstick{\text{Q2}}\\
        \lstick{\text{Q1}} & \ctrl{1} &\qswap & \targ  & \qw & & \lstick{\text{QA}}  & =\\
        \lstick{\text{QA}} & \targ & \qswap \qwx & \qw  & \qw & & \lstick{\text{Q1}}
        } & 
        \hspace{10mm}
        \Qcircuit @C=1em @R=0.9em @!R {
        \lstick{\text{Q2}} & \qw      & \qw  & \qw & \qw &\ctrl{1} & \qw & & \lstick{\text{Q2}}\\
        \lstick{\text{Q1}} & \ctrl{1} &\ctrl{1} & \targ & \ctrl{1} & \targ  & \qw & & \lstick{\text{QA}}  & =\\
        \lstick{\text{QA}} & \targ    & \targ & \ctrl{-1} & \targ & \qw  & \qw & & \lstick{\text{Q1}}
        } & 
        \hspace{10mm}
        \Qcircuit @C=1em @R=0.9em @!R {
        \lstick{\text{Q2}} & \qw & \qw &\ctrl{1} & \qw & & \lstick{\text{Q2}}\\
        \lstick{\text{Q1}} & \targ &\ctrl{1} & \targ  & \qw & & \lstick{\text{QA}}\\
        \lstick{\text{QA}} & \ctrl{-1} & \targ & \qw  & \qw & & \lstick{\text{Q1}}
        } 

    \end{tabular}
    \end{equation*}
    \caption{Example of an ssBV implementation using ancilla swapping. Here we consider the circuit for implementing the oracle with $b=11$ for ssBV-$2$ on a linear architecture with qubit connectivity Q2-Q1-QA (left). The standard swapping technique (second from left) requires $5$ CNOTs (second from right), but choosing to swap the ancilla allows the circuit to be implemented with just $3$ CNOTs (right).}
    \label{fig:bv_swaps}
\end{figure*}

\begin{figure}[t]
    \centering
    \includegraphics[trim = 0 0 0 0, clip, width=\columnwidth, valign=t]{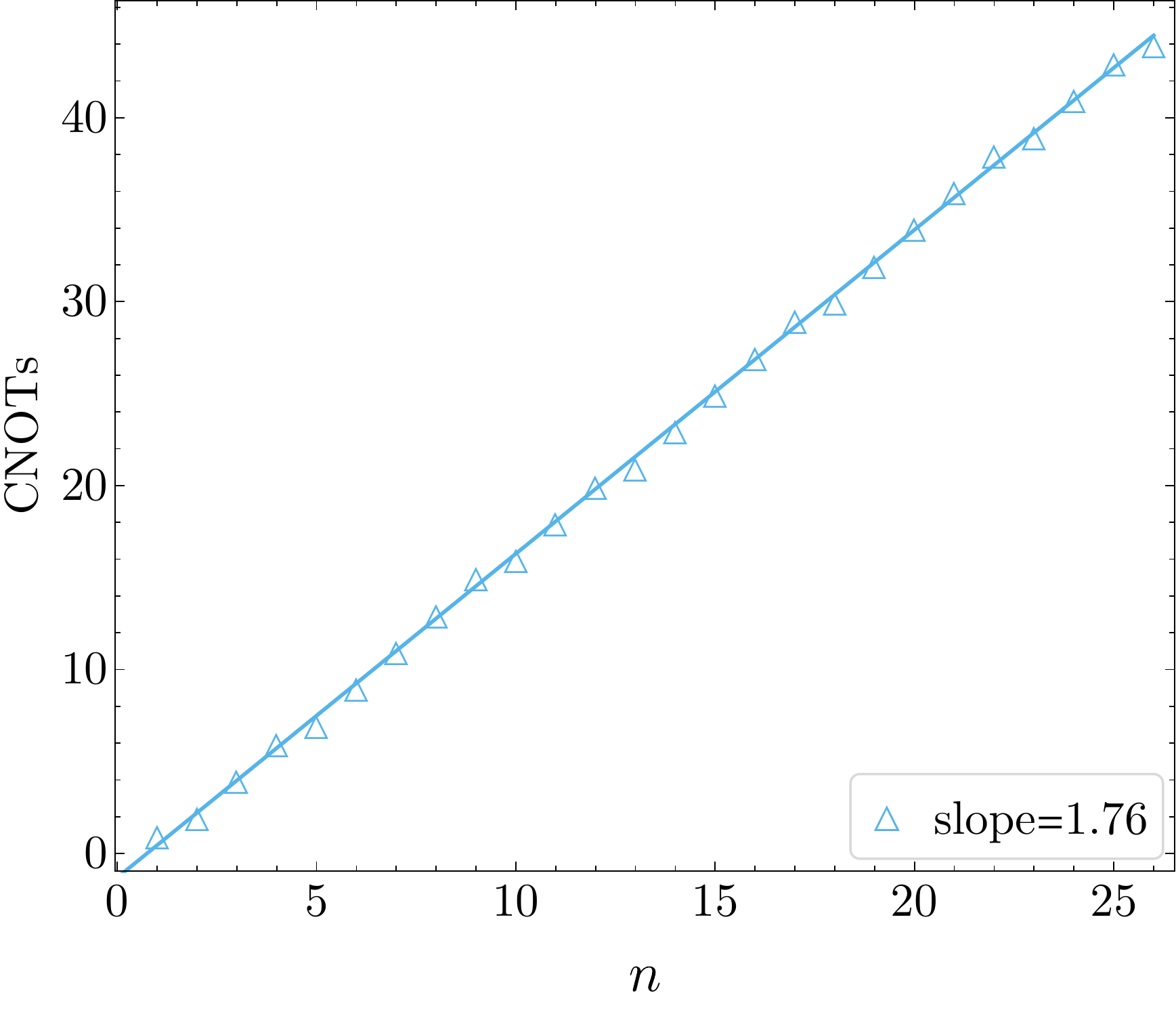} 
    \caption{The number of CNOTs used to implement the $b=1^{n}$ oracle is shown as a function of $n$. This number would scale as $2n$ for a chain, but for the heavy-hex architecture, it scales as $1.76n$. As both Montreal and Cairo use the same architecture, this number is same for both. }
     \label{fig:swap_cnot_scaling}
\end{figure}

\section{Device Specifications}
\label{app:C}

Both Montreal and Cairo are 27-qubit devices with a connectivity architecture as shown in~\cref{fig:architecture}. They have quantum volumes of 128 and 64 and are built using the IBM Quantum Falcon r4 and r5.11 processors, respectively~\cite{Pelofske:2022}. Single qubit gates are performed by driving a DRAG pulse, and two-qubit gates are echoed cross-resonance (CR) gates. $T_1$, $T_{2}$ times, and gate as well as measurement errors and durations are detailed in~\cref{tab:device}. 

\begin{table}[h]
    \centering
    \begin{tabular}{|l|c|c|c|c|c|c|}
    \hline
        ~ & \multicolumn{2}{c}{Montreal}  & & \multicolumn{2}{c}{Cairo} &  \\ \hline
        ~ & Min & Mean & Max & Min & Mean & Max \\ \hline
        $T_1$ ($\mu$s) & 57.57 & 113.2 & 187.1 & 39.93 & 102.19 & 198.93 \\ \hline
        $T_2$ ($\mu$s) & 22.58 & 99.72 & 198.71 & 18.58 & 114.19 & 290.06 \\ \hline
        1QG Error  (\%) & 0.02 & 0.04 & 0.15 & 0.01 & 0.03 & 0.07 \\ \hline
        2QG Error  (\%) & 0.57 & 1.35 & 6.09 & 0.52 & 4.64 & 100.0 \\ \hline
        1QG Duration ($\mu$s) & 0.04 & 0.04 & 0.04 & 0.02 & 0.02 & 0.02 \\ \hline
        2QG Duration ($\mu$s) & 0.27 & 0.43 & 0.63 & 0.16 & 0.31 & 0.71 \\ \hline
        RO Error (\%) & 0.79 & 2.59 & 10.66 & 0.47 & 1.39 & 4.89 \\ \hline
        RO Duration  ($\mu$s) & 5.2 & 5.2 & 5.2 & 0.73 & 0.73 & 0.73 \\ \hline
    \end{tabular}
\caption{Device specifications for Montreal and Cairo on March 12, 2022. 1QG and 2QG denote 1-qubit gate and 2-qubit gate, respectively. RO denotes readout.}
\label{tab:device}
\end{table}

Recall that in the main text we modeled the hypothetical, ideal $\text{TTS}_Q(n)$ as $c  \tau_{2q} n + \tau_0$. This linear model is actually a slight oversimplification; the ideal TTS (assuming no decoherence) for Montreal and Cairo as a function of $n$ is shown in~\cref{fig:durations_fit}, and only becomes linear for $n>2$ (Montreal) or $n>5$ (Cairo). The dashed lines shown in~\cref{fig:tts} are identical to the data shown in~\cref{fig:durations_fit} (not the linear fits). As both readout and 2-qubit gate durations are higher for Montreal, $\text{TTS}_{\text{Montreal}} > \text{TTS}_{\text{Cairo}}$. The TTS growth rates (for $n>5$) are $c \tau_{2q} =0.40 \mu$s and $c \tau_{2q} =0.27 \mu s$, and are slightly lower than the average 2-qubit gate durations of $0.43 \mu$s and $0.31 \mu$s for Montreal and Cairo, respectively. The intercepts are at $5.28\mu$s and $0.77\mu$s for Montreal and Cairo, respectively.

We used the entire chip for Montreal, but Cairo had three noisy qubits that were left out from the experiments reported in \cref{fig:tts}. To explain this, \cref{fig:cairo_faulty} shows our ssBV-$n$ results on the entire Cairo chip using a different logical-to-physical implementation than~\cref{fig:tts}. Here, the largest problem sizes solved with and without DD are $n=19$ and $25$ respectively. There is an abrupt jump in the TTS at $n=15$ due to a large readout error in Q19. In particular, the readout error for Q19 is $12.9\%$, which is an order of magnitude higher than the average readout error of $1.76\%$. Consequently, we used a different logical-to-physical embedding where the three faulty qubits, Q19 and its neighbors Q20 and Q22 (see \cref{fig:architecture}), were left out of the experiment. While this reduced the largest problem size we could solve in the DD setting (from $n=26$ to $n=23$), it allowed us to extract $\lambda$ without being affected by the anomalous readout error in Q19. Overall, we effectively treated Cairo as a $24$-qubit device in all our subsequent experiments.

\section{Reduction of circuit depth: CNOT-efficient SWAPs for ssBV-\texorpdfstring{$n$}{n}}
\label{app:D}

A fully connected architecture would allow BV-$n$ to be implemented with $n$ two-qubit gates. Both Cairo and Montreal have a maximum connectivity of $3$, and most qubits are connected to $2$ others; see~\cref{fig:architecture}. Generally, circuit transpilers deal with the sparseness by swapping qubits as necessary. The BV algorithms requires us to entangle the ``marked'' qubits (those corresponding to a $1$ in the bitstring $b$ that defines the given oracle) with the ancilla, and so we must swap the unmarked qubits when the ancilla and the marked qubit are not directly coupled. This is equivalent to swapping the ancilla instead of the marked qubit. While implementing a SWAP requires three CNOTs, by using the circuit identity $\text{CNOT}_{12} \  \text{SWAP}_{12} = \text{CNOT}_{21} \text{CNOT}_{12}$, we can implement a CNOT followed by a SWAP with just two CNOTs. \cref{fig:bv_swaps} illustrates this for the ssBV-$2$ case. Overall, swapping the ancilla reduces the CNOT scaling from $4n$ to $2n$ for a linear architecture. On the heavy-hex layout, the number of CNOTs required for implementing ssBV-$n$ is found to scale as $1.76n$, as shown in~\cref{fig:swap_cnot_scaling}. For ssBV-$26$, our longest circuit, we used $44$ CNOTs. This is because on top of the 26 CNOTs between the marked and the ancilla qubit, 18 CNOTs were needed to perform the SWAP. If each of the 18 SWAPs required 3 extra CNOTs instead, then ssBV-$26$ would require 80 CNOTs.

\section{Bootstrapping}
\label{app:E}

In all our TTS figures, we report $\text{TTS}_{\text{avg}} \equiv \text{mean}_i (\text{TTS}_i$), where $\text{TTS}_i$ is the time-to-solution for the $i$'th oracle. Each oracular experiment was repeated for 32,000 (100,000) shots on Montreal (Cairo), and we counted how many of these experiments returned the correct answer. We bootstrapped over the observed counts, using the method described in~\cite{efronBootstrapMethodsAnother1992}, to get $100$ sampled versions of each oracular experiment. The reported TTS values and the error bars correspond, respectively, to the expected value and $\pm 5\sigma$ for $\text{TTS}_{\text{avg}}$ computed from these bootstrapped samples. It is possible for the success probability for a given oracle to be non-zero and yet for some of the bootstrapped samples to have no successful counts. In such cases, we discarded those bootstrapped samples as this would lead to infinite TTS. An actual infinite TTS means that the solution was not observed during any repetition of the experiment, i.e., $p_s=0$. 

To obtain the confidence intervals for the speedup exponents $\lambda$, we first use Mathematica's LinearModelFit function to compute the worst-case fit on each bootstrapped sample. We then report the expected value with $2\sigma$ confidence intervals obtained from the ensemble of $\lambda$s.

\section{Dynamical decoupling details}
\label{app:F}

We tested a large set of known DD pulse sequences on the BV-$10$ oracle $b=1^{10}$ and then considered the performance of the top four sequences we identified in this manner on the entire ssBV-$n$ experiment (not shown). All such DD sequences outperformed the unprotected implementation on both Montreal and Cairo, and the specific choice of the DD sequence had little impact on the speedup exponents $\lambda$. The results shown in the main text utilize the universally robust ($\text{UR}_n$) sequences~\cite{genovArbitrarilyAccuratePulse2017}, which are known to perform well on superconducting devices~\cite{souzaProcessTomographyRobust2020,gautamProtectionNoisyMultipartite2021,Ezzell:22}.

The $\text{UR}_n$ sequence for $n \geq 4$ and $n$ even is defined as
\begin{subequations}
    \begin{align}
        \text{UR}_n &= (\pi)_{\phi_1} - (\pi)_{\phi_2} - \ldots - (\pi)_{\phi_n} \\
        %-------------------
        \phi_k &= \frac{(k-1)(k-2)}{2} \Phi^{(n)} + (k - 1) \phi_2 \\
        %-------------------
        \Phi^{(4m)} &= \frac{\pi}{m} \ \ \ \Phi^{(4m + 2)} = \frac{2 m \pi}{2m + 1},
    \end{align}
\end{subequations}
where $(\pi)_{\phi}$ is a $\pi$ rotation about an axis which makes an angle $\phi$ with the $x$-axis, and where $\phi_1$ is a free parameter usually set to $0$ by convention, and $\phi_2 = \pi / 2$ is a standard choice we used. This is done so that $\text{UR}_4 =$XY4, the well-known universal DD sequence~\cite{Viola:99}, as discussed in Ref.~\cite{genovArbitrarilyAccuratePulse2017}.

We implemented one repetition of the DD sequence in all available idle time gaps (see~\cref{fig:bv_circuit}) and did not attempt to optimize the pulse shape or pulse placement. Such an optimization would undoubtedly further improve performance, and presents a fruitful future research direction; see, e.g., Refs.~\cite{alex2020deep,tripathiSuppressionCrosstalkSuperconducting2022,dasADAPTMitigatingIdling2021}.

%\bibliographystyle{naturemag}
%\bibliography{refs}

\end{document}